# A new generalization of Parrondo's games to three players and its application in genetic switches

Atiyeh Fotoohinasab

August 2015

## Abstract

Parrondo's paradox indicates a paradoxical situation in which a winning expectation may occur in sequences of losing games. There are many versions of the original Parrondo's games in the literature, but the games are played by two players in all of them. We introduce a new extended version of games played by three players and a three-sided biased dice instead of two players and a biased coin in this work. In the first step, we find the part of the parameters space where the games are played fairly. After adding noise to fair probabilities, we combine two games randomly, periodically, and nonlinearly and obtain the conditions under which the paradox can occur. This generalized model can be applied in all science and engineering fields. It can also be used for genetic switches. Genetic switches are often made by two reactive elements, but the existence of more elements can lead to more existing decisions for cells. Each genetic switch can be considered a game in which the reactive elements compete to increase their molecular concentrations. We present three genetic networks based on a new generalized Parrondo's games model, consisting of two noisy genetic switches. The combination of them can increase network robustness to noise. Each switch can also be used as an initial pattern to construct a synthetic switch to change undesirable cells' fate.

## 1. Introduction

Brownian ratchet thought experiment in physics is about periodic and asymmetric moves of Brownian particles to the right in the effect of periodic or random arrangement of switching on and off the potential. The Brownian ratchet mechanism's full description is explained in the Feynman Lectures on Physics [1]. The Parrondo's paradox is a discrete-time version of the Brownian ratchet's physical model, and they are related via the Fokker–Planck equation [2-6].

The original Parrondo's paradox, which was discovered by Parrondo in 1960, is a contradictory situation that two individually losing games, A and B, can produce a winning expectation when they are combined randomly or periodically [7-9]. After each coin-flipping, the player's capital will increase (decrease) by one unit if a head (tail) event is observed. Game A is a biased coin-tossing game (coin one) with the winning probability $p = \frac{1}{2} - \epsilon$ and the losing probability $q = \frac{1}{2} + \epsilon$. On the other hand, game B is played by tossing two biased coins, coin two and coin three, which the winning probabilities of them are $p_2 = \frac{1}{10} - \epsilon$ and $p_3 = \frac{3}{4} - \epsilon$, respectively. If the player one's capital is multiple of three, then coin two is tossed; otherwise, coin three is tossed. Player one will lose in both individual games in the long run but playing these games in random or periodic order can produce a winning expectation. It has also been proven that the paradox will happen in chaotic switching between games [10, 11].

In Parrondo's games, game B has a critical role in the occurrence of a paradoxical situation. Convex probability space is a necessary condition for the occurrence of paradox so that a combination of two losing games can shift into an area with a winning expectation [9, 12, 13]. Some extensions of the original Parrondo's games have been proposed [14-16]. All of them are divided into three general categories: capital-dependent [5, 7, 17], history-dependent [18, 19] and spatial neighbor-dependent [20-22]. The classical game B is an example of a capital-dependent game that the winning probabilities of game B depend on the player's capital value. In the history-dependent version, the winning probabilities of game B depend on the recent history of winning and losing games. For the first two categories, game A is the same as the original game. The last version of games, spatial neighbor-dependent, so-called cooperative, is played by more than one player. According to the periodic boundary conditions assumption, players are arranged in a circle, and each of them possesses a capital $C_i(t), i = 1,2,..,N$ (for $N$ players). In the cooperative version, game A has its original setup. A biased coin is flipped repeatedly while winning probabilities of game B depend on the winning or losing the state of neighbors of the player who is chosen to play. We can consider the left and right neighbors of the present player (one-dimensional case) [22] or his left, right, up, and below neighbors (two-dimensional case) [20]. Many structures of game B were also designed with arbitrary topologies in which the number of the neighbors of the nodes is not consistent [23-25]. There are many extended versions of Parrondo's games in these three categories, but all of them have been played by two players. In this article, we present a new extension of Parrondo's games in which the two games are played by three players and rolling a three-sided dice instead of two players and coin-flipping.

Many examples of counterintuitive dynamics can be observed in several fields, ranging from biology to economics, such as control theory, theory of granular flow, diffusion processes, population genetics, and physical quantum systems [3, 14, 26-33]. Some studies have also been presented about the application of Parrondo's paradox in biology. In biological systems, it may be related to gene transcription dynamics in GCN4 protein and the dynamics of transcription errors in DNA [34]. Reed [35] claimed that a two-locus system under selection could be constructed like Parrondo's games. Bier [26] presented the essential role of Brownian motion in motor proteins' action (individual molecules that convert chemical energy into motion).

In this work, we introduce a new application of Parrondo's paradox in genetic switches. A genetic switch consists of a number of genes that can inhibit or activate its own or others gene expression by their productions [36-38]. These genetic switches are usually composed of two elements, but the presence of more than two switch elements can produce multiple (more than two) stable states. For example, we can mention the GATA-PU.1 genetic switch in the cellular differentiation process of hematopoietic stem cells [37] and the generation of four cell types in Bacillus subtilis [39]. It can be inferred from the result of our generalized model that genetic switches with more than two elements (three elements in our model) in combination together can also enhance the network resistance to noise [40, 41]. Noise plays a pivotal role in genetic switches. In biological systems, it is usually considered to be an adverse effect. However, the role of noise can be deleterious or useful related to the situations where the system is under them [42, 43]. Noise can lead to cell survival in stressful environments by randomly changing the cell state to a more compatible state. Conversely, the cell state may be switched to an undesirable state [44-46]. Here, we introduce three networks of games in which the games can be modeled as a genetic switch with three reactive elements. These reactive elements are the players of the game that compete together for increasing their concentrations. The transition probabilities to each of the system's states are adjusted according to the environment conditions [47]; for this purpose, the tuning parameter is the biasing parameter in our model. If one of the two switches exits from its natural path by entered noise, another switch can return it to its correct path.

In section two, we first briefly review the original Parrondo's paradox. In section three, our new strategy for generalization of Parrondo's games to three players is introduced. In this section, the requirements for having fair, losing, and winning games are also presented by some theoretical studies. We combine two games, A and B, in a randomized, periodical, and nonlinear manner and prove that the paradox cannot occur in that strategy of games for randomized combination. In section four, three networks of games will be represented, which can model some genetic switches in biology. These models have been inspired by possible relations between genes.

## 2. Orginal Parrondo's Parradox

Parrondo shows that combining two losing strategies of games can lead to a winning result if we periodically or randomly switch between them [7, 8, 48]. The original version of Parrondo's games consists of two players and two fair gambling games, games A and game B. Both of them are losing games after applying a positive biasing parameter $\varepsilon$. Indeed, there are conditions under which one of the players loses when plays two games individually and wins in a combined game counterintuitively. Hereinafter, we deal with this player. The structures of Parrondo's games are represented diagrammatically in Figure 1.

Each time a coin is tossed, one of the games is played and time increases by one unit, and players' capital increases (decreases) by one, if he wins (loses). Harmer et al. in [7] have detailed the governing equations of the two games theoretically. Here, we describe Parrondo's games briefly and express ultimate equations.

- **Game A:** This game is a biased coin flipping game. The probability of occurrence of a head event (winning probability) is $p$, and the tail event (losing probability) is $q = 1 - p$. It is obvious that game A is a losing game for $p < 0.5$.
- **Game B:** In this game, there are two biased coins, coin two and coin three, where their winning probabilities are different. If the player's current capital is a multiple of $M$, coin two with winning probability $p_1$ is tossed; otherwise, coin three with winning probability $p_2$ is tossed. This game is winning, losing, or fair when the inequality $\frac{(1-p_1)(1-p_2)^M}{p_1 p_2^M}$ is less than 1, greater than one, or equal to 1.
- **Randomized combined game:** consider game A is played with probability $\gamma$ and game B with probability $1 - \gamma$ in each round of the combined game, then the probability that the combined game is won is $\acute{p}_1 = \gamma p + (1-\gamma)p_1$ if the current capital is a multiple of $M$; whereas if the current capital is not a multiple of $M$, the winning probability of this game is $\acute{p}_2 = \gamma p + (1-\gamma)p_2$. The losing probabilities are $\acute{q}_1 = 1 - \acute{p}_1$ and $\acute{q}_2 = 1 - \acute{p}_2$, respectively. It is evident that these probabilities are identical to game $B$, except that their values are different. So, according to the respective equation for game B, it can be inferred that the randomized game is winning, losing, and fair if $\frac{(1-\acute{p}_1)(1-\acute{p}_2)^M}{\acute{p}_1 \acute{p}_2{}^M}$ is less than 1, greater than one or equal to 1.

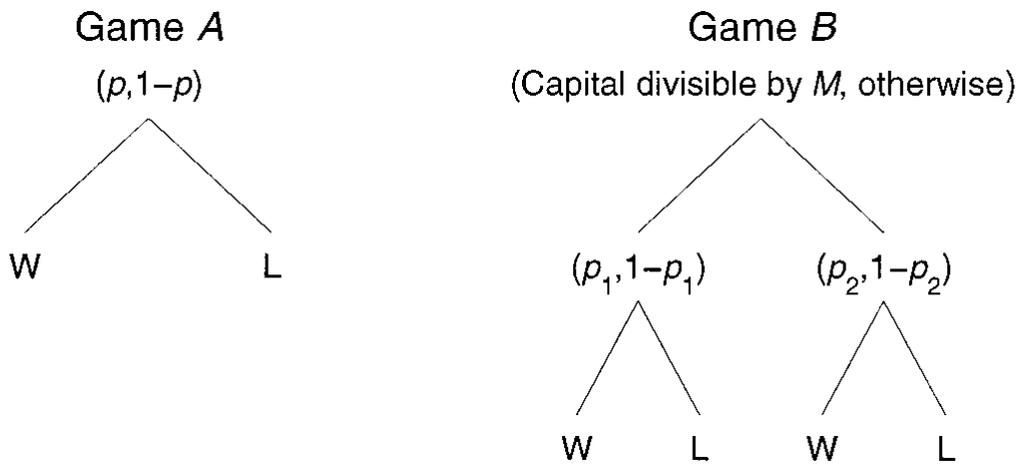

Figure 1. Construction of Parrondo's games. Game B is a capital-dependent game [7]

If $M = 3$ and $\gamma = \frac{1}{2}$ are chosen, the game B is fair for any solutions in the following set of parameters space:

$$0 < p_1 < \frac{1}{2} \quad and \quad p_2 = \frac{-1 + p_1}{-1 + 2p_1} - \sqrt{\frac{p_1 - p_1{}^2}{(-1 + 2p_1)^2}}$$

Or

$$\frac{1}{2} < p_1 < 1 \quad and \quad p_2 = \frac{-1 + p_1}{-1 + 2p_1} + \sqrt{\frac{p_1 - p_1{}^2}{(-1 + 2p_1)^2}}$$

In the original game B, the probabilities $p_1 = \frac{1}{10}$ and $p_2 = \frac{3}{4}$ are selected from this set. After adding noise to the fair probabilities, they are $p = \frac{1}{2} - \epsilon$, $p_1 = \frac{1}{10} - \epsilon$, $p_2 = \frac{3}{4} - \epsilon$ for both the games A and B. The games are fair if $\epsilon = 0$ and are losing if $\epsilon > 0$. It has been proven that selecting any biasing parameter in the range of $0 < \epsilon < 0.01311$ can lead to a paradoxical situation. It is also shown that combining two games periodically and nonlinearly can return the winning game [10, 11].

## 3. Generalization of Parrondo's games to three players

These new games are based on competition of three players instead of two players in original games; therefore, they play with rolling a three-sided dice instead of flipping a coin. Whenever a dice is thrown, one of the games is played, and time increases by one unit. Each player's capital increases by two units or decreases by one unit, given that he is a winner or loser, respectively. In fact, the winner receives two dollars from the other two players. We now describe the details of each game separately and then combine two games using randomized, periodic, and nonlinear methods.

### 3.1. Game A

Game A is a simple game and is equivalent to a biased random walk on a triangle in the plane. We call the three players player one, player two, and player three. Like the original game, we associate each side of the dice to a player, so the outcome of dice rolls specifies the winner; if the outcome of the throw is one, player one is the winner (this is also true for other players). We have summarized the rules of game A as follows:

i) If the dice roll outcome is one, player one is a winner, and his capital increases by two units. We characterize this event's probability by $p_1^A$ (the winning probability of player one in game A).

ii) If the outcome of the dice roll is two, player two is a winner and his capital increases by two units. We characterize this event's probability by $p_2^A$ (the winning probability of player two in game A).

iii) If the outcome of the dice roll is three, player three is a winner and his capital increases by two units. We characterize this event's probability by $p_3^A$ (the winning probability of player three in game A).

The above rules can be expressed mathematically in this way:

$$Game\ A = \begin{cases} p(the\ winning\ probability\ of\ player\ one) = p_1^A \\ p(the\ winning\ probability\ of\ player\ two) = p_2^A \\ p(the\ winning\ probability\ of\ player\ three) = p_3^A \end{cases} \quad (1)$$

that $p_1^A + p_2^A + p_3^A = 1$.

We seek a situation where both fair games A and B with a biasing tunable parameter $\varepsilon$ are individually losing games. Their combination is a winning game in any alternating order. For this purpose, we first obtain the possible probabilities in which the game A is fair, and then we find the conditions that the player one will lose in the long run. Assuming $f_j^{\ 1}$ characterizes the probability that the player one's capital reaches zero in a finite number of games, given that we start with capital $j$. According to Markov chain theory [49], player one in this game is either;

i) If $f_j^{\ 1} = 1, f_j^{\ 2} = 1,$ and $f_j^{\ 3} = 1$ for all $j \geq 0$, the game will be fair.

ii) If $f_j^{\ 1} = 1, f_j^{\ 2} < 1$ or $f_j^{\ 3} < 1$ for all $j > 0$, player one will lose.

iii) If $f_j^{\ 1} < 1, f_j^{\ 2} = 1,$ and $f_j^{\ 3} = 1$ for all $j > 0$, player one will win.

We can find $f_j^{\ 1}$ by the following recursive equation:

$$f_j^{\ 1} = p_1^A f_{j+2}^{1} + (1 - p_1^A) f_{j-1}^{1} \quad (2)$$

for $j \geq 0$ and subject to $f_0^{\ 1} = 1$.

Same as player one, there are similar conditions for player two and three. The recursive equations for these players are as follows:

$$f_j^{\ 2} = p_2^A f_{j+2}^{2} + (1 - p_2^A) f_{j-1}^{2} \quad (3)$$

$$f_j^3 = p_3^A f_{j+2}^3 + (1 - p_3^A) f_{j-1}^3 \tag{4}$$

for $j \geq 0$ and the initial conditions $f_0^2 = 1$ and $f_0^3 = 1$. In the above equations, $f_j^2$ and $f_j^3$ are the probabilities that the capital of player two and player three reaches zero in a finite number of games, given that they start with capital $j$.

If we solve the recursive relations (2)-(4) concerning the condition (i), we obtain the following solutions for having a fair game:

$$\frac{-p_1^A + \sqrt{4p_1^A - 3(p_1^A)^2}}{2p_1^A} = 1, \quad i.e\ p_1^A = \frac{1}{3} \tag{5a}$$

$$\frac{-p_2^A + \sqrt{4p_2^A - 3(p_2^A)^2}}{2p_2^A} = 1, \quad i.e\ p_2^A = \frac{1}{3} \tag{5b}$$

$$\frac{-p_3^A + \sqrt{4p_3^A - 3(p_3^A)^2}}{2p_3^A} = 1, \quad i.e\ p_3^A = \frac{1}{3} \tag{5c}$$

Also according to condition (ii), player one will be lost in the long run when below inequalities hold:

$$\frac{-p_1^A + \sqrt{4p_1^A - 3(p_1^A)^2}}{2p_1^A} > 1, \quad i.e\ p_1^A < \frac{1}{3} \tag{6a}$$

$$\frac{-p_2^A + \sqrt{4p_2^A - 3(p_2^A)^2}}{2p_2^A} < 1, \quad i.e\ p_2^A > \frac{1}{3} \tag{6b}$$

$$\frac{-p_3^A + \sqrt{4p_3^A - 3(p_3^A)^2}}{2p_3^A} < 1, \quad i.e\ p_3^A > \frac{1}{3} \tag{6c}$$

### 3.2. Game B

Here, the structure of game B is based on a state-dependent category. Like the original game B, the probabilities of winning for each player on each round of the game are conditional and dependent on the player's capital (losing player). The strategy of this game can be considered mathematically as follows:

$$Game\ B = \begin{cases} p\begin{pmatrix} \text{the winning probability} \\ \text{of player one} \end{pmatrix} \begin{vmatrix} \text{the capital of player one is} \\ \text{a multiple of } M \end{vmatrix} = p_{1,1}^B \\ p\begin{pmatrix} \text{the winning probability} \\ \text{of player one} \end{pmatrix} \begin{vmatrix} \text{the capital of player one} \\ \text{is not a multiple of } M \end{vmatrix} = p_{2,1}^B \\ p\begin{pmatrix} \text{the winning probability} \\ \text{of player two} \end{pmatrix} \begin{vmatrix} \text{the capital of player one} \\ \text{is a multiple of } M \end{vmatrix} = p_{1,2}^B \\ p\begin{pmatrix} \text{the winning probability} \\ \text{of player two} \end{pmatrix} \begin{vmatrix} \text{the capital of player one} \\ \text{is not a multiple of } M \end{vmatrix} = p_{2,2}^B \\ p\begin{pmatrix} \text{the winning probability} \\ \text{of player there} \end{pmatrix} \begin{vmatrix} \text{the capital of player one} \\ \text{is a multiple of } M \end{vmatrix} = p_{1,3}^B \\ p\begin{pmatrix} \text{the winning probability} \\ \text{of player three} \end{pmatrix} \begin{vmatrix} \text{the capital of player one} \\ \text{is not a multiple of } M \end{vmatrix} = p_{2,3}^B \end{cases} \quad (7)$$

It is obvious that $p_{1,1}^B + p_{1,2}^B + p_{1,3}^B = 1$ and $p_{2,1}^B + p_{2,2}^B + p_{2,3}^B = 1$.

Like the original game B, $M=3$ is chosen. Let $g_j^k$ for $k \in \{1,2,3\}$ be the probability that the capital of player k ever reaches zero, given that he starts with $j$ units. From Markov chain theory, player one is either;

i) If $g_j^1 = 1, g_j^2 = 1,$ and $g_j^3 = 1$ for all $j \geq 0$, the game will be fair.

ii) If $g_j^1 > 1, g_j^2 < 1,$ and $g_j^3 < 1$ for all $j > 0$, player one will lose.

iii) If $g_j^1 < 1, g_j^2 > 1,$ and $g_j^3 > 1$ for all $j > 0$, player one will win.

Again following the derivation of equations of game A, for $i \geq 0$ and $j \in \{1,2\}$, the set of numbers $\{g_k^1\}$ is found by the following recursive equation:

$$g_{3i}^1 = p_{1,1}^B g_{3i+2}^1 + (1 - p_{1,1}^B) g_{3i-1}^1 \quad (8a)$$

And

$$g_{3i+j}^1 = p_{2,1}^B g_{3i+j+2}^1 + (1 - p_{2,1}^B) g_{3i+j-1}^1 \quad (8b)$$

subject to the boundary condition $g_0^1 = 1$. According to the condition i, after some tedious manipulation and for $i \geq 1$, the following difference equation is written by using the above recursive relations:

$$(p_{1,1}^B \cdot (p_{2,1}^B)^2) g_{3i+6}^1 + \left(2 \cdot p_{1,1}^B \cdot p_{2,1}^B \cdot (1 - p_{2,1}^B) + (p_{2,1}^B)^2 \cdot (1 - p_{1,1}^B)\right) g_{3i+3}^1 +$$
$$\left((p_{1,1}^B \cdot (1 - p_{2,1}^B)^2) + (2 \cdot (1 - p_{1,1}^B) \cdot (1 - p_{2,1}^B) \cdot p_{2,1}^B) - 1\right) g_{3i}^1 +$$
$$\left((1 - p_{1,1}^B) \cdot (1 - p_{2,1}^B)^2\right) g_{3i-3}^1 = 0 \tag{9}$$

There is also a similar difference equation for player two:

$$(p_{1,2}^B \cdot (p_{2,2}^B)^2) g_{3i+6}^2 + \left(2 \cdot p_{1,2}^B \cdot p_{2,2}^B \cdot (1 - p_{2,2}^B) + (p_{2,2}^B)^2 \cdot (1 - p_{1,2}^B)\right) g_{3i+3}^2 +$$
$$\left((p_{1,2}^B \cdot (1 - p_{2,2}^B)^2) + (2 \cdot (1 - p_{1,2}^B) \cdot (1 - p_{2,2}^B) \cdot p_{2,2}^B) - 1\right) g_{3i}^2 +$$
$$\left((1 - p_{1,2}^B) \cdot (1 - p_{2,2}^B)^2\right) g_{3i-3}^2 = 0 \tag{10}$$

subject to the boundary condition $g_0^2 = 1$. This equation is identical to Eq. (9) except that its coefficients have changed, so they have similar solutions. The probability spaces, in which the game B is played fairly, is constructed from the common area of solutions of these equations. The following set shows the possible probabilities for having a fair game:

$$\left\{ 0 < p_{1,1}^B < 1, \quad p_{2,1}^B = \frac{1 - p_{1,1}^B}{2}, \right.$$
$$0 < p_{1,2}^B < 1 - p_{1,1}^B, \quad p_{2,2}^B = \frac{1 - p_{1,2}^B}{2}, \tag{11}$$
$$\left. p_{1,3}^B = 1 - p_{1,1}^B - p_{1,2}^B, \quad p_{2,3}^B = 1 - p_{2,1}^B - p_{2,2}^B \right\}$$

and the player one will also lose in long run if the following inequalities hold:

$$\frac{-2p_{1,1}^B p_{2,1}^B - (p_{2,1}^B)^2 + 2p_{1,1}^B (p_{2,1}^B)^2 + \sqrt{4p_{1,1}^B (p_{2,1}^B)^2 - 4p_{1,1}^B (p_{2,1}^B)^3 + (p_{2,1}^B)^4}}{2p_{1,1}^B (p_{2,1}^B)^2} > 1 \tag{12a}$$

$$\frac{-2p_{1,2}^B p_{2,2}^B - (p_{2,2}^B)^2 + 2p_{1,2}^B (p_{2,2}^B)^2 + \sqrt{4p_{1,2}^B (p_{2,2}^B)^2 - 4p_{1,2}^B (p_{2,2}^B)^3 + (p_{2,2}^B)^4}}{2p_{1,2}^B (p_{2,2}^B)^2} < 1 \tag{12b}$$

$$\frac{-2p_{1,3}^B p_{2,3}^B - (p_{2,3}^B)^2 + 2p_{1,3}^B (p_{2,3}^B)^2 + \sqrt{4p_{1,3}^B (p_{2,3}^B)^2 - 4p_{1,3}^B (p_{2,3}^B)^3 + (p_{2,3}^B)^4}}{2p_{1,3}^B (p_{2,3}^B)^2} < 1 \tag{12c}$$

In the next three sections, we examine the existence of paradox for combination of these new generalizations of games in random, periodic and non-linear orders.

### 3.3. Randomized Switching

Now we are describing the combined game in randomized strategy. Consider the situation in which game A is played with probability $\gamma$ and game B is played with probability $1-\gamma$. We will show this combined game as $C$. In the following, there is the mathematical form of the combined game:

$$Game\ C = \begin{cases} p\begin{pmatrix} the\ winning\ probability \\ of\ player\ one \end{pmatrix}\begin{vmatrix} the\ capital\ of\ player\ one \\ is\ a\ multiple\ of\ M \end{vmatrix} = p_{1,1}{}^C \\ p\begin{pmatrix} the\ winning\ probability \\ of\ player\ one \end{pmatrix}\begin{vmatrix} the\ capital\ of\ player\ one \\ is\ not\ a\ multiple\ of\ M \end{vmatrix} = p_{2,1}{}^C \\ p\begin{pmatrix} the\ winning\ probability \\ of\ player\ two \end{pmatrix}\begin{vmatrix} the\ capital\ of\ player\ one \\ is\ a\ multiple\ of\ M \end{vmatrix} = p_{1,2}{}^C \\ p\begin{pmatrix} the\ winning\ probability \\ of\ player\ two \end{pmatrix}\begin{vmatrix} the\ capital\ of\ player\ one \\ is\ not\ a\ multiple\ of\ M \end{vmatrix} = p_{2,2}{}^C \\ p\begin{pmatrix} the\ winning\ probability \\ of\ player\ three \end{pmatrix}\begin{vmatrix} the\ capital\ of\ player\ one \\ is\ a\ multiple\ of\ M \end{vmatrix} = p_{1,3}{}^C \\ p\begin{pmatrix} the\ winning\ probability \\ of\ player\ three \end{pmatrix}\begin{vmatrix} the\ capital\ of\ player\ one \\ is\ not\ a\ multiple\ of\ M \end{vmatrix} = p_{2,3}{}^C \end{cases} \quad (13)$$

Which the probabilities will be:

$$p_{1,1}^C = \gamma p_1^A + (1-\gamma)p_{1,1}^B$$

$$p_{2,1}^C = \gamma p_1^A + (1-\gamma)p_{2,1}^B$$

$$p_{1,2}^C = \gamma p_2^A + (1-\gamma)p_{1,2}^B \quad (14)$$

$$p_{2,2}^C = \gamma p_2^A + (1-\gamma)p_{2,2}^B$$

$$p_{1,3}^C = \gamma p_3^A + (1-\gamma)p_{1,3}^B$$

$$p_{2,3}^C = \gamma p_3^A + (1-\gamma)p_{2,3}^B$$

Like the original game, the combined game form is identical to game B, but its probabilities have been modified. Therefore, player one will win in the long run in the combined game if the below inequalities hold:

$$\frac{-2p_{1,1}^C p_{2,1}^C - (p_{2,1}^C)^2 + 2p_{1,1}^C (p_{2,1}^C)^2 + \sqrt{4p_{1,1}^C (p_{2,1}^{A+B})^2 - 4p_{1,1}^C (p_{2,1}^C)^3 + (p_{2,1}^C)^4}}{2p_{1,1}^C (p_{2,1}^C)^2} < 1 \quad (15a)$$

$$\frac{-2p^C_{1,2}p^C_{2,2}-(p^C_{2,2})^2+2p^C_{1,2}(p^C_{2,2})^2+\sqrt{4p^C_{1,2}(p^C_{2,2})^2-4p^C_{1,2}(p^C_{2,2})^3+(p^C_{2,2})^4}}{2p^C_{1,2}(p^C_{2,2})^2} > 1 \qquad (15b)$$

$$\frac{-2p^C_{1,3}p^C_{2,3}-(p^C_{2,3})^2+2p^C_{1,3}(p^C_{2,3})^2+\sqrt{4p^C_{1,3}(p^C_{2,3})^2-4p^C_{1,3}(p^C_{2,3})^3+(p^C_{2,3})^4}}{2p^C_{1,3}(p^C_{2,3})^2} > 1 \qquad (15c)$$

Now we add the tunable noise parameters $\varepsilon_1$ and $\varepsilon_2$ to the fair probabilities of game A and B, as mentioned in Eq. (11) and (5a)-(5c):

$$S = \begin{Bmatrix} p^A_1 - \varepsilon_1, p^A_2 - \varepsilon_2, p^A_3 + \varepsilon_1 + \varepsilon_2; \; p^B_{1,1} - \varepsilon_1, p^B_{1,2} - \varepsilon_2, p^B_{1,3} + \varepsilon_1 + \varepsilon_2, \\ p^B_{2,1} - \varepsilon_1, \; p^B_{2,2} - \varepsilon_2, p^B_{2,3} + \varepsilon_1 + \varepsilon_2 \end{Bmatrix} \qquad (16)$$

We substitute these probabilities into equations (6), (12), and (14), then using the conclusive equation (14), rewrite the Eq. (15a)-(15c) and take intersection between the final bounds of equations (6), (12) and (15) for losing conditions of game A and B and winning conditions of the combined game and the fair probabilities of game A and B in Eq. 11 and Eq. 15 respect to $\varepsilon > 0$. Unfortunately, they have nothing in common in the parameter space, and the paradox cannot occur in the randomized strategy in our generalized model. However, we have shown that paradox can obtain when the two games combine periodically and nonlinearly by some simulation and for some parameters.

### 3.4. Periodic Switching

In this section, we investigate the effect of combining two losing games A and B, periodically in our generalized model by adding positive noise $\varepsilon_1$ and $\varepsilon_2$ to the fair probabilities of games A and B. Our simulations were performed so on until 100 games were played, which were averaged over 10000 trials, and the biasing parameters were chosen as $\varepsilon_1 = 0.005$ and $\varepsilon_2 = 0.001$. Therefore the probabilities of game A are $p^A_1 = \frac{1}{3} - \varepsilon_1, p^A_2 = \frac{1}{3} - \varepsilon_2$, and $p^A_3 = \frac{1}{3} + \varepsilon_1 + \varepsilon_2$, which according to the selected positive noises, this game will lose. We examine the existence of paradox for different sequences of games A and B in the combined game and different probabilities of game B. According to Eq.(11), consider the two below subsets of fair probabilities of game B, $S_1$, and $S_2$:

$$S_1 = \{p^B_{1,1} = 0.1, \; p^B_{1,2} = 0.2, p^B_{1,3} = 0.7; \; p^B_{2,1} = 0.45, \; p^B_{2,2} = 0.4, \; p^B_{2,3} = 0.15\}$$

$$S_2 = \{p^B_{1,1} = 0.25, p^B_{1,2} = 0.3, p^B_{1,3} = 0.45;\ p^B_{2,1} = 0.375, p^B_{2,2} = 0.35, p^B_{2,3} = 0.275\}$$

Figures 2 and 3 show the results of playing individual games A and B and the periodic combination of them for two different sequences of games, before and after adding noise parameters $\varepsilon_1 = 0.005$ and $\varepsilon_2 = 0.001$ to the fair possibilities subsets $S_1$ and $S_2$ (in accordance with relation 16). These figures demonstrate that player one will lose in the long run of playing individual games A and B under the determinate noise parameters $\varepsilon_1 = 0.005$ and $\varepsilon_2 = 0.001$, but combining these two losing games together by sequences with the order of AABAAB… and AABBBBAABBBB... produces a winning expectation in his gain. According to Figures 2f and 2h (also Figures 3f and 3h), the rate of winning for sequences of games with a long period is low. The efficacy of $p^B_{1,1}$ in determining the probabilities parameters of game B in Eq.13 is dominant, so even though there are many different sequences of games that could lead to the occurrence of paradox, in all of them, the player one in the combined game is the winner until the probability around $p^B_{1,1} = 0.3$, and then the game will be a fair game for a small range value of $p^B_{1,1}$ and finally, he will lose for the greater value of $p^B_{1,1}$. These results are valid for sequences that start with game A. Figure 4 is plotted for the sequence BBABBA... with the fair probabilities $p^B_{1,1} = 0.6, p^B_{1,2} = 0.3, p^B_{1,3} = 0.1;\ p^B_{2,1} = 0.2, p^B_{2,2} = 0.35, p^B_{2,3} = 0.45$ and noise parameters $\varepsilon_1 = 0.005$ and $\varepsilon_2 = 0.001$ which starts with game B. It is noteworthy that when we combine two fair games A and B, there is a winning expectation for player one, same as the original game. In the future, we represent a new strategy for game B, in which the two losing games A and B combine together and produce a fair game, whereas the combined game without adding noise is also fair.

### 3.5. Chaotic switching

In the original Parrondo's paradox, the games A and B are combined based on randomly or periodically choosing one of the two games. However, Tang et al. [10, 11] use a new strategy in switching between the two games based on various chaotic time series through simulation results. These numeric sequences are generated by different chaotic systems, which strongly depend on the initial conditions. They are often quite similar to random sequences in the time domain, so they are plotted in the phase space. In this way, the chaotic sequence can be easily identified because there is a regular pattern in the phase space plot.

There are many one-dimensional and two-dimensional chaotic maps like Logistic map [50], Sinusoidal map [51], Tent map [51] and Gaussian map [51, 52] in a one-dimensional dynamic and the Henon map [51] and Lozi [52] map in two-dimensional dynamic. We use this method in our work and test the existence of paradox for the Logistic map chaotic time series sequence generator. It is one of the most straightforward dynamic systems showing chaotic behavior, and its equation is:

- Logistic Map

$$x_{n+1} = ax_n(1 - x_n)$$

The coefficient of this chaotic generator will determine the stability of it. It is shown that under the stable regions, the system behaves periodically. According to Tang et al. [10], the maximum rate of winning for Parrondo's games occurs when the chaotic generator tends toward periodic behavior. For choosing the system's coefficient, we use the bifurcation diagram of the generator and select the value under which the system is stable (for this value, the sequence is periodic). There are many ways to switch between the games based on this strategy, but the easiest and most used way is to compare each value of a chaotic sequence with an appropriate constant $\gamma$.

We choose $a = 3.85$, $x_0 = 0.1$ and $\gamma = 0.5$ for our simulations. It was previously pointed that chaotic time series sequence generators depend on the initial conditions, so there will be different solutions. Figures 5 and 6 show the result of combining two games nonlinearly for two subsets $S_1$ and $S_2$ of fair probabilities of game B. As can be seen, player one will lose in two games, but he will win by combining these games.

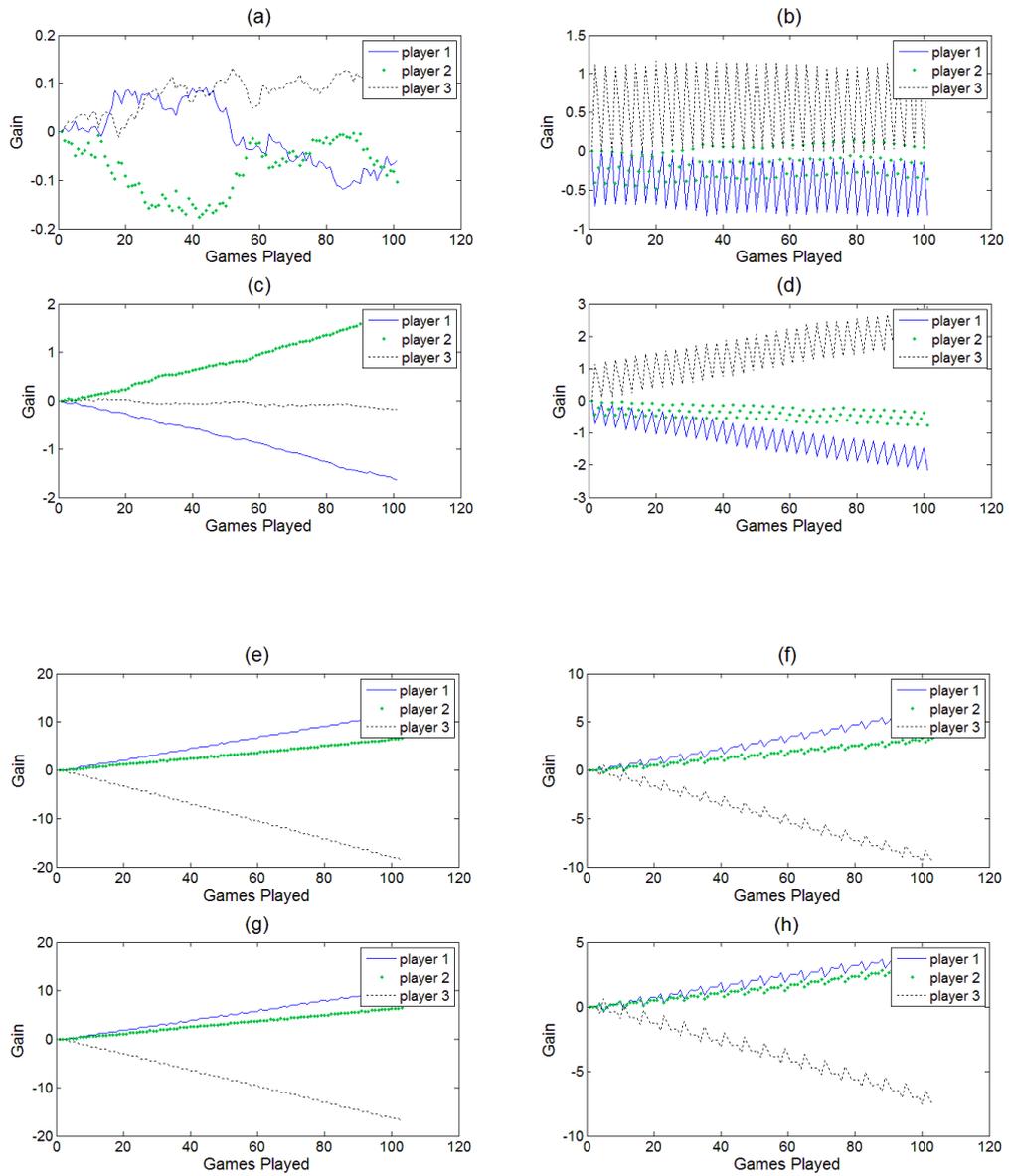

Figure 2. Individual games and a combination of them before and after adding noise to the fair probabilities $S_1$. a) the effect of playing game A individually before adding noise, b) the effect of playing game B individually before adding noise, c) the effect of noise on game A, d) the effect of noise on game B, e) the effect of periodic switching between two fair games A and B in order of sequence AABAAB…, f) the effect of periodic switching between two noisy games A and B in order of sequence AABAAB…, g) the effect of periodic switching between two fair games A and B in order of sequence AABBBBAABBBB…, h) the effect of periodic switching between two noisy games A and B in order of sequence AABBBBAABBBB….

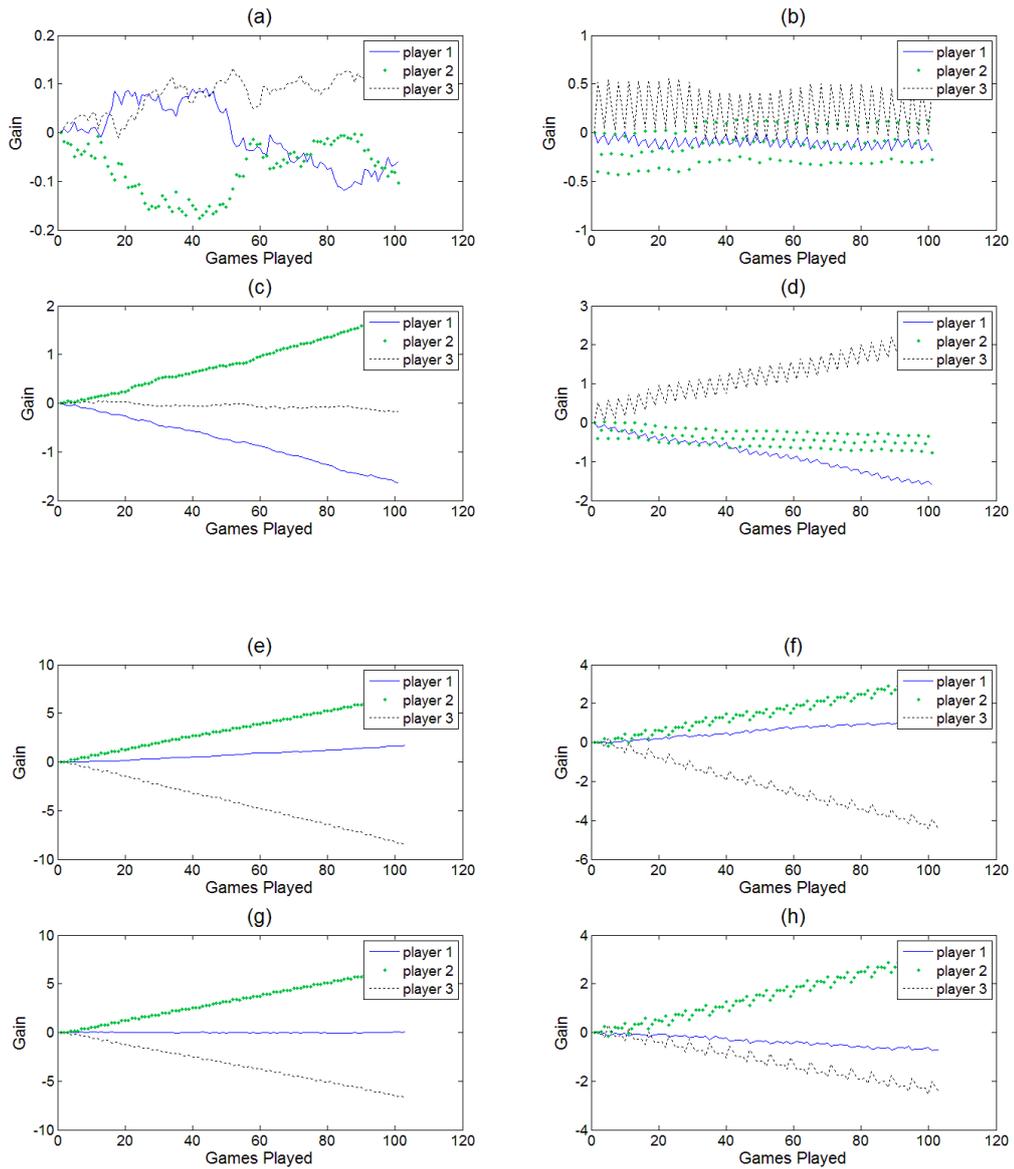

Figure 3. Individual games and a combination of them before and after adding noise to the fair probabilities $S_2$. a) the effect of playing game A individually before adding noise, b) the effect of playing game B individually before adding noise, c) the effect of noise on game A, d) the effect of noise on game B, e) the effect of periodic switching between two fair games A and B in order of sequence AABAAB…, f) the effect of periodic switching between two noisy games A and B in order of sequence AABAAB…, g) the effect of periodic switching between two fair games A and B in order of sequence AABBBBAABBBB…, h) the effect of periodic switching between two noisy games A and B in order of sequence AABBBBAABBBB….

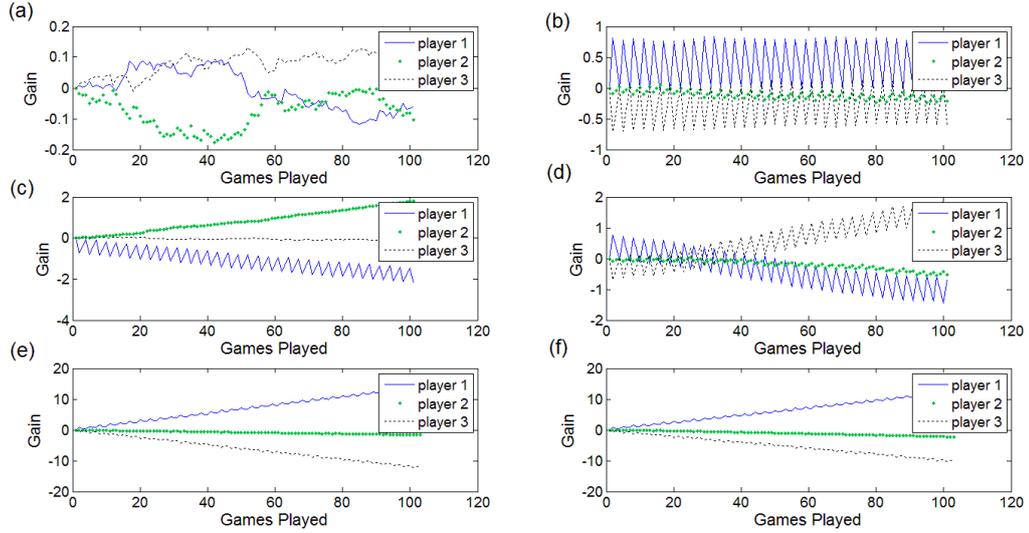

*Figure 4.* Individual games and combination of them before and after adding noise to the fair probabilities in set $\left\{p_1^A = \frac{1}{3},\ p_2^A = \frac{1}{3}, p_3^A = \frac{1}{3};\ p_{1,1}^B = 0.6,\ p_{1,2}^B = 0.3, p_{1,3}^B = 0.1;\ p_{2,1}^B = 0.2, p_{2,2}^B = 0.35,\ p_{2,3}^B = 0.45\right\}$. a) the effect of playing game A individually before adding noise, b) the effect of playing game B individually before adding noise, c) the effect of noise on game A, d) the effect of noise on game B, e) the effect of periodic switching between two fair games A and B in order of sequence BBABBA…, f) the effect of periodic switching between two noisy games A and B in order of sequence BBABBA ….

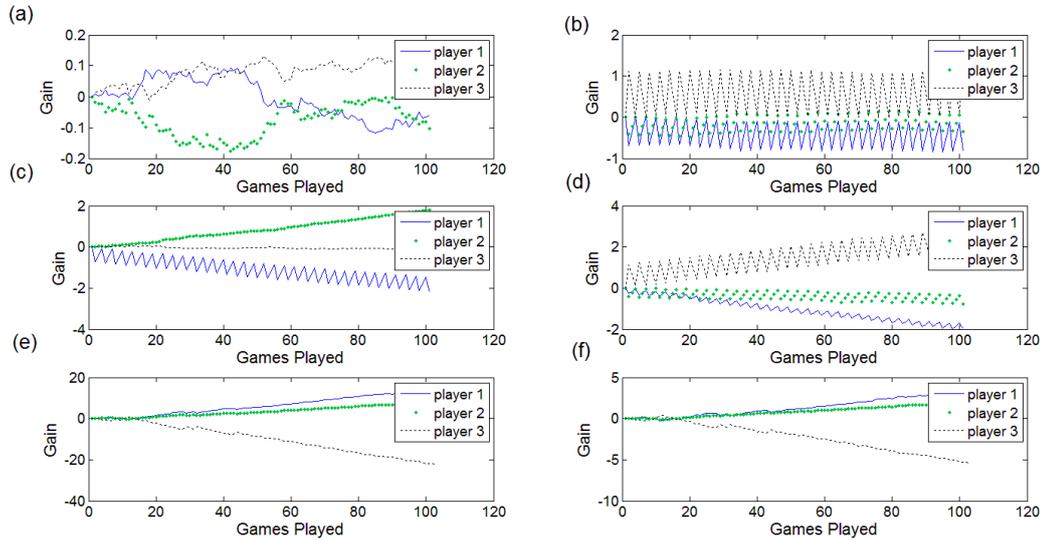

*Figure 5. Nonlinear combination of individual games before and after adding noise to the fair possibilities set $S_1$.* a) the effect of playing game A individually before adding noise, b) the effect of playing game B individually before adding noise, c) the effect of noise on game A, d) the effect of noise on game B, e) the effect of nonlinear switching between two fair games A and B, f) the effect of nonlinear switching between two noisy games A and B.

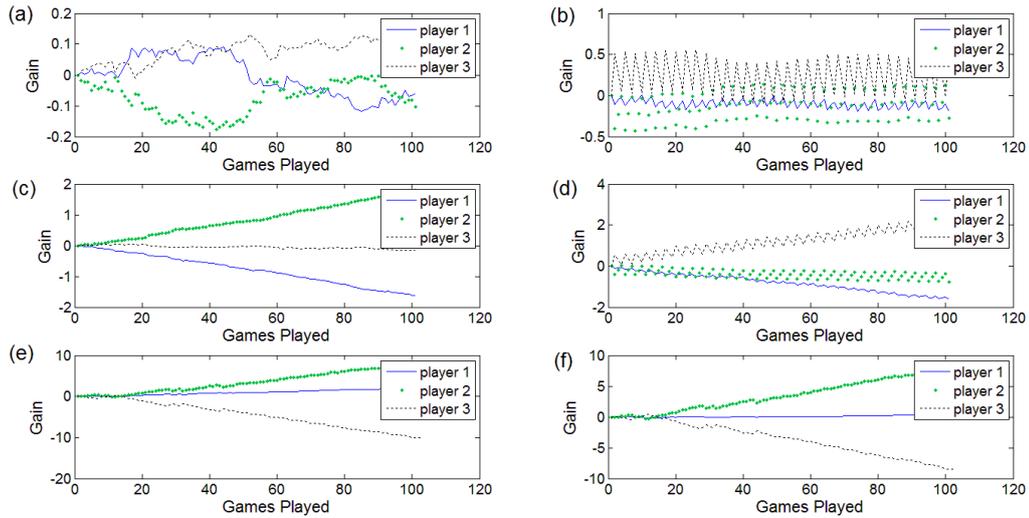

*Figure 6. Nonlinear combination of individual games before and after adding noise to the fair possibilities set $S_2$. a) the effect of playing game A individually before adding noise, b) the effect of playing game B individually before adding noise, c) the effect of noise on game A, d) the effect of noise on game B, e) the effect of nonlinear switching between two fair games A and B, f) the effect of nonlinear switching between two noisy games A and B.*

## 4. Games based on genetic switches

The described generalized model can be applied in several fields, such as two-player games. One of its applications can be in genetic switches [53]. The generalized model is inappropriate for biological networks because of game B's structure and the nonexistence of paradox in randomized switching. In contrast, in biological networks, decision making occurs randomly. In the next sections, we present three networks of games that can be fitted to genetic switches. The determinate conditions for games specify the requirements for feedback loops in genetic switches equivalent to these games [54-57]. In these networks, game A models a simplex switch, and game B models a conditional switch. These games can be diverse regulatory motifs in cells that the combination of them can reduce the destructive effects of noise, or each strategy of switches can be used to design a synthetic switch to change the cell fate [40, 58-61].

Our used game theoretic approach for modeling genetic switches is a kind of the Stochastic Simulation Algorithm (SSA) [46, 62], in which a reaction is fired per unit of time, depending on the current state of the system. We consider a system of molecules with three chemical species $\{S_1, S_2, S_3\}$ as the players in genetic switches where they compete against each other to increase their molecular concentrations. These chemical species interact through $M$ chemical reactions $R_1, R_2, \ldots, R_M$. The number of all reactions that can fire in the system is equal to triple of the number of conditions in a game. The system can only randomly choose one of the three possible reactions per unit of time (the possibility of winning one of the three players in each round of the games). Because switch A is a simple switch, the three reactions are constant in each state. We present a three-dimensional discrete state vector $X(t) = \{X_1(t), X_2(t), X_3(t)\}$ where its integer values $X_i(t)$ represent the number of molecules of species $S_i$ at time t. Every time a reaction is fired, the number of molecules of one of the chemical species increases two units, and other species decrease one unit in consonance with their winning or losing situation. A stoichiometric vector $v_j = \{v_{1,j}, v_{2,j}, v_{3,j}\}$ is associated to reaction $R_j$ which elements $v_{i,j}$ represents the change in $X_i$ after firing reaction $R_j$. These stoichiometric vectors form columns of the stoichiometry matrix $D = \{v_1\ v_2\ ...\ v_M\}$. Thus, a system in the state $X(t) = x$ goes to the new state $x + v_j$ after firing reaction $R_j$. The probability that a reaction $R_j$ fire in a particular state of the system is equivalent to winning probabilities of players in each game's condition. We summarize our game-theoretic approach for modeling switch B in algorithm 1. This algorithm computes the realization of state vector $X(t)$, starting from state $x_0$ at time $t_0$ and up to time $T$. The used algorithm for modeling switch A is similar to algorithm 1 except that it is not necessary to do step 3 when this switch is without any condition. We use an analogous algorithm for combining these biological switches as a combined game, except that after the second step, we should choose one of the switches (games) from a Bernoulli distribution.

### *Algorithm* 1 *SSA* $(t_0, x_0, T)$

1: *set* $x \leftarrow x_0 = 0$ *and* $t \leftarrow t_0 = 0$;

2: *While* $t < T = 100$ *do*

3: *according to the* concentration of special species, choose one of the possible states of the system;

4: *according to the state of the system, choose j as a Generalized Bernoulli random variable distribution from the three possibilities of reactions*;

5: *set* $x \leftarrow x + v_j$ *and* $t \leftarrow t + 1$;

6: *end while*

For each reactive element, we can assign one of the following statuses;

*Fair status*: If the rate of increase in the number of molecules of a reactive element is zero, this element is in the fair status. In other words, the number of molecules of this reactive element remains around its initial number.

*Winning status*: If the rate of increase in the number of molecules of a reactive element is positive, this element is in the winning status.

*Losing status*: If the rate of increase in the number of molecules of a reactive element is negative, this element is in the losing status.

Finally, due to the status of three reactive elements, a cell chooses a fate. Now, we describe our networks in the next three sections.

### 4.1. Network I

In this network, game A is a simple switch with three reactive elements that compete against each other without any condition. Similar to game A in the generalized model, the fair probabilities for three elements in switch A are $p_1^A = \frac{1}{3}$, $p_2^A = \frac{1}{3}$, and $p_3^A = \frac{1}{3}$. The strategy of this switch can be considered mathematically as follow:

$$switch\ A = \begin{cases} p\begin{pmatrix} \text{the probability of increasing the} \\ \text{number of molecules of the} \\ \text{element one by two unit at each step} \end{pmatrix} = p_1^A \\ p\begin{pmatrix} \text{the probability of increasing the} \\ \text{number of molecules of the} \\ \text{element two by two unit at each step} \end{pmatrix} = p_2^A \\ p\begin{pmatrix} \text{the probability of increasing the} \\ \text{number of molecules of the} \\ \text{element three by two unit at each step} \end{pmatrix} = p_3^A \end{cases} \qquad (17)$$

Our strategy for switch B has two states and has been inspired by a fictitious three-gene cellular switch where one of the genes (e.g., gene one) determines the cell fate. Due to the competition result between switch elements, the number of molecules produced by each gene increases by two units or decreases by one. We define $C_1$ to characterize the number of molecules of the reactive elements one and $t^B$ to specify the required threshold for changing the system's state. This switch can be described mathematically as follow:

$$switch\ B = \begin{cases} p\begin{pmatrix} \text{the probability of increasing the} \\ \text{number of molecules of the} \\ \text{element one by two unit at each step} \end{pmatrix} \Big| C_1 \geq t^B \end{pmatrix} = p_{1,1}{}^B \\ p\begin{pmatrix} \text{the probability of increasing the} \\ \text{number of molecules of the} \\ \text{element one by two unit at each step} \end{pmatrix} \Big| C_1 < t^B \end{pmatrix} = p_{2,1}{}^B \\ p\begin{pmatrix} \text{the probability of increasing the} \\ \text{number of molecules of the} \\ \text{element two by two unit at each step} \end{pmatrix} \Big| C_1 \geq t^B \end{pmatrix} = p_{1,2}{}^B \\ p\begin{pmatrix} \text{the probability of increasing the} \\ \text{number of molecules of the} \\ \text{element two by two unit at each step} \end{pmatrix} \Big| C_1 < t^B \end{pmatrix} = p_{2,2}{}^B \\ p\begin{pmatrix} \text{the probability of increasing the} \\ \text{number of molecules of the} \\ \text{element three by two unit at each step} \end{pmatrix} \Big| C_1 \geq t^B \end{pmatrix} = p_{1,3}{}^B \\ p\begin{pmatrix} \text{the probability of increasing the} \\ \text{number of molecules of the} \\ \text{element three by two unit at each step} \end{pmatrix} \Big| C_1 < t^B \end{pmatrix} = p_{2,3}{}^B \end{cases} \quad (18)$$

Using simulation and subject to $t^B = 0$, we choose the following solution for switches A and B in which the three reactive elements are in the fair status:

$$S_3 = \begin{cases} p_1{}^A = \frac{1}{3}, p_2{}^A = \frac{1}{3}, p_3{}^A = \frac{1}{3}; \\ p_{1,2}^B = 0.1, p_{1,2}^B = 0.41, p_{1,3}^B = 0.49; \\ p_{2,1}^B = 0.71, p_{2,2}^B = 0.21, p_{2,3}^B = 0.08 \end{cases}$$

In a random order combination of the switches, if switch A is used with probability $\gamma$ and switch B with probability $1 - \gamma$, until $\gamma = 0.6$ the element one will be in fair status, and afterward, the number of its molecules will increase over time (winning status). We add positive noise parameters $\varepsilon_1$ and $\varepsilon_2$ to fair probabilities in the set of $S_3$ (see Eq.16). Simulations have been plotted for two different choices of two parameters $\varepsilon_1$ and $\varepsilon_2$ in Figures 7 and 8. These figures show changes in the number of molecules over time steps. The simulation results are performed so on until 100 steps were done, and each step is averaged over 10000 repeats. As shown in Figures 7c and 7e, the number of molecules of element one in switch A has decreased over time. Given to various values of $\gamma$ in Figure 8, the number of molecules of element one in the combination case can be in fair or winning status. However, in all of these cases, element one is in the fair status in the switch B. Using the switch A with high probability (for large amounts of $\gamma$) in which the element one is in the losing status, element one has a winning expectation in the number of its molecules in the combination case.

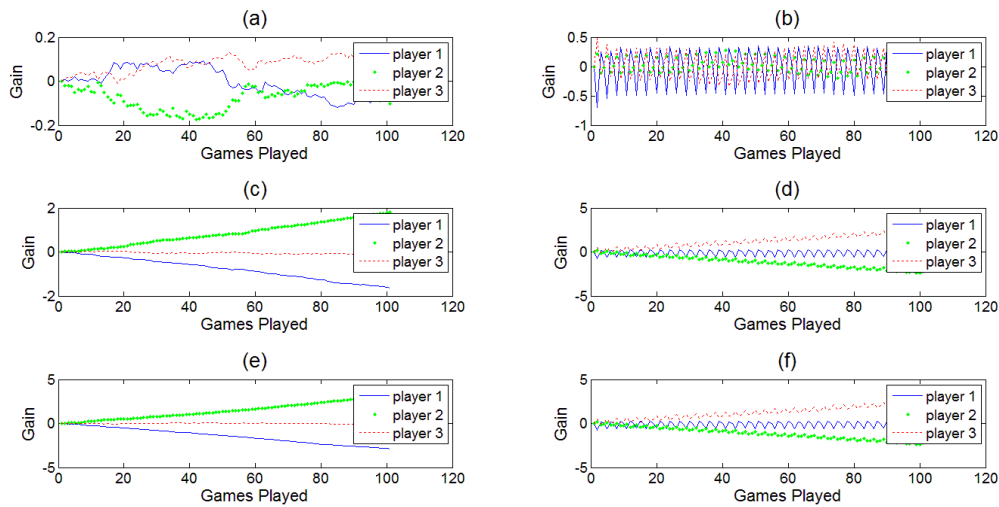

Figure 7. Individual switches for various amounts of noise parameters in network I. a) switch A before adding noise, b) switch B before adding noise, c) the effect of noise parameters $\varepsilon_1 = 0.005$ and $\varepsilon_2 = 0.001$ on switch A, d) the effect of noise parameters $\varepsilon_1 = 0.005$ and $\varepsilon_2 = 0.001$ on switch B, e) the effect of noise parameters $\varepsilon_1 = 0.01$ and $\varepsilon_2 = 0.005$ on switch A, f) the effect of noise parameters $\varepsilon_1 = 0.01$ and $\varepsilon_2 = 0.005$ on switch B.

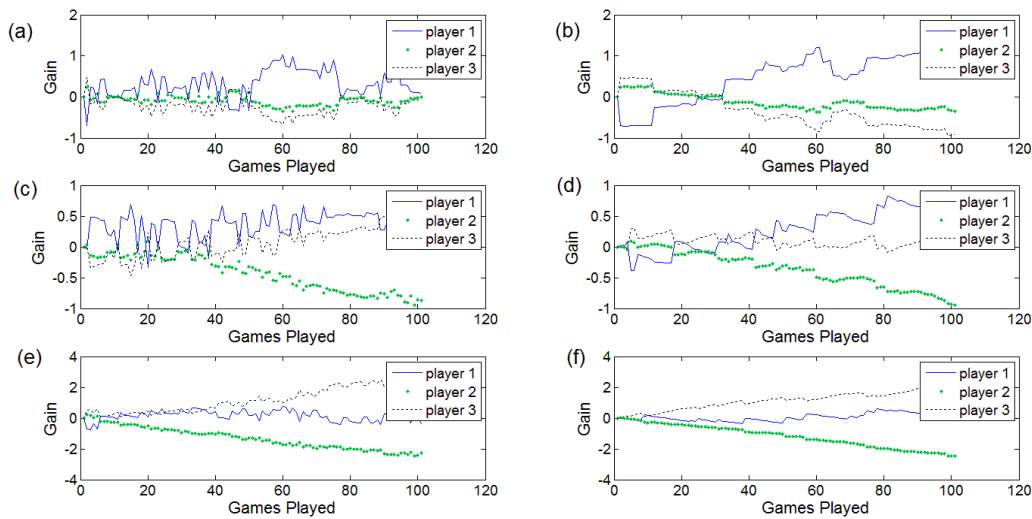

Figure 8. A random combination of individual switches for various amounts of noise parameters and switching thresholds network I. a) the effect of randomized switching with switching threshold $\gamma = 0.5$ between two fair switches A and B, b) the effect of randomized switching with switching threshold $\gamma = 0.8$ between two fair switches A and B, c) the effect of randomized switching with switching threshold $\gamma = 0.5$ between two noisy switches A and B which infected by noise parameters $\varepsilon_1 = 0.005$ and $\varepsilon_2 = 0.001$, d) the effect of randomized switching with switching threshold $\gamma = 0.8$ between two noisy switches A and B which infected by noise parameters $\varepsilon_1 = 0.005$ and $\varepsilon_2 = 0.001$, e) the effect of randomized switching with switching threshold $\gamma = 0.5$ between two noisy switches A and B which infected by noise parameters $\varepsilon_1 = 0.01$ and $\varepsilon_2 = 0.005$, f) the effect of randomized switching with switching threshold $\gamma = 0.8$ between two noisy switches A and B which infected by noise parameters $\varepsilon_1 = 0.01$ and $\varepsilon_2 = 0.005$.

### 4.2. Network II

In this network, switch A is a simple switch and has the strategy of switch A in network I with fair probabilities $p_1^A = \frac{1}{3}$, $p_2^A = \frac{1}{3}$, and $p_3^A = \frac{1}{3}$ (see Relation 17). Switch B is a conditional switch that has three states. We assume that the number of molecules of reactive elements one and two determines the state of switch B. As before, the number of winning element molecules increases two units, and the other ones decrease one unit. If we characterize $C_1$ and $C_2$ as the number of molecules of reactive elements one and two respectively, the strategy of switch B is described mathematically as relation 19. Using simulation, we found the following set of parameters for switches A and B in which the three reactive elements are in the fair status:

$$S_4 = \begin{cases} p_1^A = \frac{1}{3}, p_2^A = \frac{1}{3}, p_3^A = \frac{1}{3}; \\ p_{1,1}^B = 0.7, \ p_{1,2}^B = 0.2 \ , p_{1,3}^B = 0.1; \\ p_{2,1}^B = 0.21, \ p_{2,2}^B = 0.51, \ p_{2,3}^B = 0.28; \\ p_{3,1}^B = 0.15, \ p_{3,2}^B = 0.20, \ p_{3,3}^B = 0.65 \end{cases}$$

$$\text{switch } B = \begin{cases} p\begin{pmatrix}\text{the probability of increasing}\\\text{the number of molecules}\\\text{of the element one}\\\text{by two units at each step}\end{pmatrix} \middle| C_1 \leq 0) = p_{1,1}^B \\ p\begin{pmatrix}\text{the probability of increasing}\\\text{the number of molecules}\\\text{of the element one}\\\text{by two units at each step}\end{pmatrix} \middle| C_1 > 0 \text{ and } C_2 \leq 0) = p_{2,1}^B \\ p\begin{pmatrix}\text{the probability of increasing}\\\text{the number of molecules}\\\text{of the element one}\\\text{by two units at each step}\end{pmatrix} \middle| C_1 > 0 \text{ and } C_2 > 0) = p_{3,1}^B \\ p\begin{pmatrix}\text{the probability of increasing}\\\text{the number of molecules}\\\text{of the element two}\\\text{by two units at each step}\end{pmatrix} \middle| C_1 \leq 0) = p_{1,2}^B \\ p\begin{pmatrix}\text{the probability of increasing}\\\text{the number of molecules}\\\text{of the element two}\\\text{by two units at each step}\end{pmatrix} \middle| C_1 > 0 \text{ and } C_2 \leq 0) = p_{2,2}^B \\ p\begin{pmatrix}\text{the probability of increasing}\\\text{the number of molecules}\\\text{of the element two}\\\text{by two units at each step}\end{pmatrix} \middle| C_1 > 0 \text{ and } C_2 > 0) = p_{3,2}^B \\ p\begin{pmatrix}\text{the probability of increasing}\\\text{the number of molecules}\\\text{of the element three}\\\text{by two units at each step}\end{pmatrix} \middle| C_1 \leq 0) = p_{1,3}^B \\ p\begin{pmatrix}\text{the probability of increasing}\\\text{the number of molecules}\\\text{of the element three}\\\text{by two units at each step}\end{pmatrix} \middle| C_1 > 0 \text{ and } C_2 \leq 0) = p_{2,3}^B \\ p\begin{pmatrix}\text{the probability of increasing}\\\text{the number of molecules}\\\text{of the element three}\\\text{by two units at each step}\end{pmatrix} \middle| C_1 > 0 \text{ and } C_2 > 0) = p_{3,3}^B \end{cases} \quad (19)$$

The changes in the number of molecules of each reactive element along the time in individual switches and random combination of them due to the two various sets of noise parameters $\varepsilon_1$ and $\varepsilon_2$ are shown in Figures 9 and 10. To have better exposure, we plotted the results so on until 200 time steps were gone forward and then averaged over 10000 repeats. For the two chosen sets of noise parameters, element one in switch A is in the losing status, and in switch B, it is in the fair status (see Figure 9). As before, we use switch A with probability $\gamma$ and switch B with probability $1 - \gamma$ in each time step. By choosing $\gamma = 0.5$, the number of molecules of element one is increased slightly in the primary steps, but after that, its rate of increases will be zero in the long term. In the primary steps, the rate of increases is slower for noisy cases than when the noise parameters are $\varepsilon_1 = 0$ and $\varepsilon_2 = 0$. Using switch A, in which the decreasing rate for the number of molecules of element one is high, with high probability ($\gamma = 0.8$) produces an increasing expectation in the number of molecules of element one over time (see Figure 10).

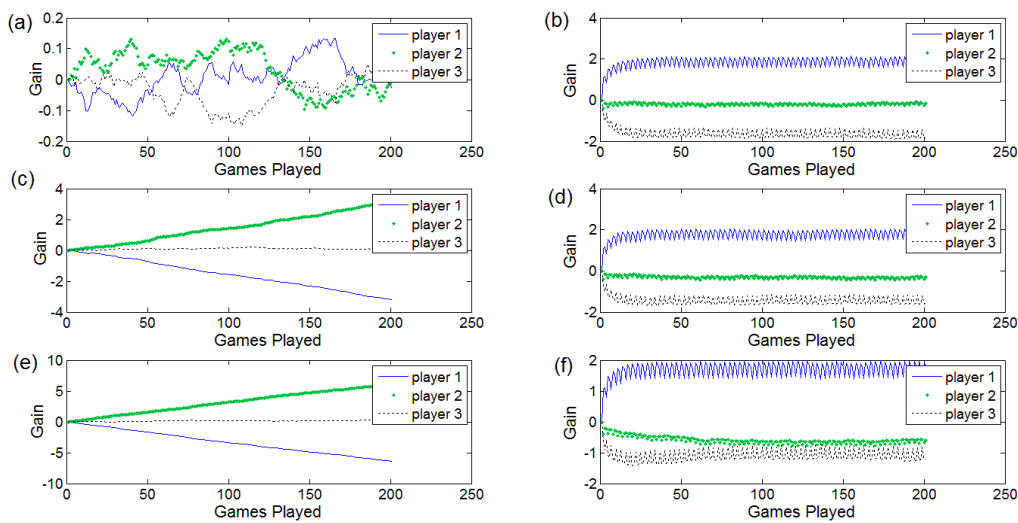

*Figure 9.* Individual switches for various amounts of noise parameters in network II. a) switch A before adding noise, b) switch B before adding noise, c) the effect of noise parameters $\varepsilon_1 = 0.005$ and $\varepsilon_2 = 0.001$ on switch A, d) the effect of noise parameters $\varepsilon_1 = 0.005$ and $\varepsilon_2 = 0.001$ on switch B, e) the effect of noise parameters $\varepsilon_1 = 0.01$ and $\varepsilon_2 = 0.005$ on switch A, f) the effect of noise parameters $\varepsilon_1 = 0.01$ and $\varepsilon_2 = 0.005$ on switch B.

### 4.3. Network III

This network is equivalent to the generalized model except that we separate cases with remainder one and two in game B when $M = 3$. Switch A has a simplex form similar to the previous networks, and its fair probabilities are $p_1^A = \frac{1}{3}$, $p_2^A = \frac{1}{3}$, and $p_3^A = \frac{1}{3}$ (see Relation 17). The switch B strategy has been inspired by a fictitious three-gene genetic switch where one gene (e.g., gene one) determines the cell fate. The number of molecules produced by each gene increases by two units or decreases by one. Each gene's productions can be bound to the upstream sequences of itself or other genes to activate or inhibit the gene expression of itself or others. Assume that every three molecules tend to bind together to produce a trimmer protein.

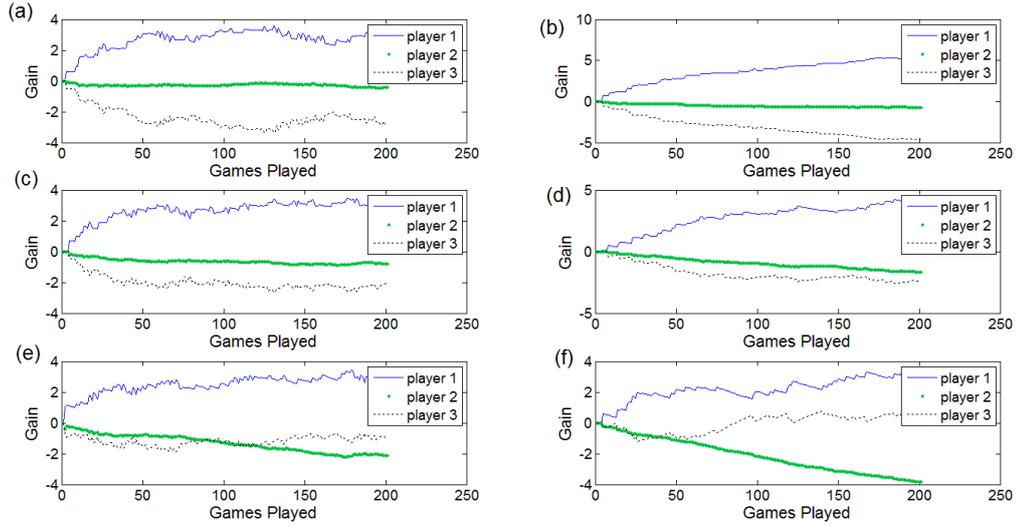

*Figure 10.* A random combination of individual switches for various amounts of noise parameters and switching thresholds in network II. a) the effect of randomized switching with switching threshold $\gamma = 0.5$ between two fair switches A and B, b) the effect of randomized switching with switching threshold $\gamma = 0.8$ between two fair switches A and B, c) the effect of randomized switching with switching threshold $\gamma = 0.5$ between two noisy switches A and B which infected by noise parameters $\varepsilon_1 = 0.005$ and $\varepsilon_2 = 0.001$, d) the effect of randomized switching with switching threshold $\gamma = 0.8$ between two noisy switches A and B which infected by noise parameters $\varepsilon_1 = 0.005$ and $\varepsilon_2 = 0.001$, e) the effect of randomized switching with switching threshold $\gamma = 0.5$ between two noisy switches A and B which infected by noise parameters $\varepsilon_1 = 0.01$ and $\varepsilon_2 = 0.005$, f) the effect of randomized switching with switching threshold $\gamma = 0.8$ between two noisy switches A and B which infected by noise parameters $\varepsilon_1 = 0.01$ and $\varepsilon_2 = 0.005$.

Depend on the number of protein molecules produced by gene one; the cell is in one of the following states at each time step:

i) There are just some ternary complexes from molecules of reactive element one.

ii) There are some ternary complexes from molecules of reactive element one with a single molecule.

iii) There are some ternary complexes from molecules of reactive element one with two single molecules.

For the second situation, we define a new probability parameter $\mu$ that describes a single molecule's tendency to promoters of the other two genes. Considering these descriptions, the strategy of switch B can be considered mathematically as relation 20. The theoretical solution for obtaining the fair probabilities of reactive elements in switch B is equivalent to which we applied in the generalized model section. Figure 11 shows the result of combination switches A and B for the fair probabilities in set $S_5$. The simulation results are performed so on until 100 steps were done, and each step is averaged over 10000 repeats. This strategy, different from the others and even the original Parrondo's paradox, produces a fair expectation from two noisy switches in which the combination of two fair switches is fair.

$$switch\ B = \begin{cases}
p\begin{pmatrix} \text{the probability of increasing} \\ \text{the number of molecules} \\ \text{of the element one, by} \\ \text{two units at each step} \end{pmatrix} \begin{vmatrix} \text{there are just some ternary} \\ \text{complexes from molecules of} \\ \text{reactive element one} \end{vmatrix} = p^B_{1,1} \\
p\begin{pmatrix} \text{the probability of increasing} \\ \text{the number of molecules} \\ \text{of the element one, by} \\ \text{two units at each step} \end{pmatrix} \begin{vmatrix} \text{there are some ternary} \\ \text{complexes from molecules} \\ \text{of reactive element one} \\ \text{with a single molecule and } \mu \leq t \end{vmatrix} = p^B_{2,1} \\
p\begin{pmatrix} \text{the probability of increasing} \\ \text{the number of molecules} \\ \text{of the element one, by} \\ \text{two units at each step} \end{pmatrix} \begin{vmatrix} \text{there are some ternary} \\ \text{complexes from molecules} \\ \text{of reactive element one} \\ \text{with a single molecule and } \mu > t \end{vmatrix} = p^B_{3,1} \\
p\begin{pmatrix} \text{the probability of increasing} \\ \text{the number of molecules} \\ \text{of the element one, by} \\ \text{two units at each step} \end{pmatrix} \begin{vmatrix} \text{there are some ternary} \\ \text{complexes from molecules} \\ \text{of reactive element one} \\ \text{with two single molecules} \end{vmatrix} = p^B_{4,1} \\
p\begin{pmatrix} \text{the probability of increasing} \\ \text{the number of molecules} \\ \text{of the element two, by} \\ \text{two units at each step} \end{pmatrix} \begin{vmatrix} \text{there are just some ternary} \\ \text{complexes from molecules of} \\ \text{reactive element one} \end{vmatrix} = p^B_{1,2} \\
p\begin{pmatrix} \text{the probability of increasing} \\ \text{the number of molecules} \\ \text{of the element two, by} \\ \text{two units at each step} \end{pmatrix} \begin{vmatrix} \text{there are some ternary} \\ \text{complexes from molecules} \\ \text{of reactive element one} \\ \text{with a single molecule and } \mu \leq t \end{vmatrix} = p^B_{2,2} \\
p\begin{pmatrix} \text{the probability of increasing} \\ \text{the number of molecules} \\ \text{of the element two by} \\ \text{two units at each step} \end{pmatrix} \begin{vmatrix} \text{there are some ternary} \\ \text{complexes from molecules} \\ \text{of reactive element one} \\ \text{with a single molecule and } \mu > t \end{vmatrix} = p^B_{3,2} \\
p\begin{pmatrix} \text{the probability of increasing} \\ \text{the number of molecules} \\ \text{of the element two by} \\ \text{two units at each step} \end{pmatrix} \begin{vmatrix} \text{there are some ternary} \\ \text{complexes from molecules} \\ \text{of reactive element one} \\ \text{with two single molecules} \end{vmatrix} = p^B_{4,2} \\
p\begin{pmatrix} \text{the probability of increasing} \\ \text{the number of molecules} \\ \text{of the element three by} \\ \text{two units at each step} \end{pmatrix} \begin{vmatrix} \text{there are just some ternary} \\ \text{complexes from molecules of} \\ \text{reactive element one} \end{vmatrix} = p^B_{1,3} \\
p\begin{pmatrix} \text{the probability of increasing} \\ \text{the number of molecules} \\ \text{of the element three by} \\ \text{two units at each step} \end{pmatrix} \begin{vmatrix} \text{there are some ternary} \\ \text{complexes from molecules} \\ \text{of reactive element one} \\ \text{with a single molecule and } \mu \leq t \end{vmatrix} = p^B_{2,3} \\
p\begin{pmatrix} \text{the probability of increasing} \\ \text{the number of molecules} \\ \text{of the element three by} \\ \text{two units at each step} \end{pmatrix} \begin{vmatrix} \text{there are some ternary} \\ \text{complexes from molecules} \\ \text{of reactive element one} \\ \text{with a single molecule and } \mu > t \end{vmatrix} = p^B_{3,3} \\
p\begin{pmatrix} \text{the probability of increasing} \\ \text{the number of molecules} \\ \text{of the element three by} \\ \text{two units at each step} \end{pmatrix} \begin{vmatrix} \text{there are some ternary} \\ \text{complexes from molecules} \\ \text{of reactive element one} \\ \text{with two single molecules} \end{vmatrix} = p^B_{4,3}
\end{cases} \quad (20)$$

$$S_5 = \begin{cases} p_1^A = \frac{1}{3}, p_2^A = \frac{1}{3}, p_3^A = \frac{1}{3}; \\ p_{1,1}^B = 0.5, p_{1,2}^B = 0.3, p_{1,3}^B = 0.2; \\ p_{2,1}^B = 0.3, p_{2,2}^B = 0.2, p_{2,3}^B = 0.5; \\ p_{3,1}^B = 0.5, p_{3,2}^B = 0.4, p_{3,3}^B = 0.1; \\ p_{4,1}^B = 0.1, p_{4,2}^B = 0.4, p_{4,3}^B = 0.5 \end{cases}$$

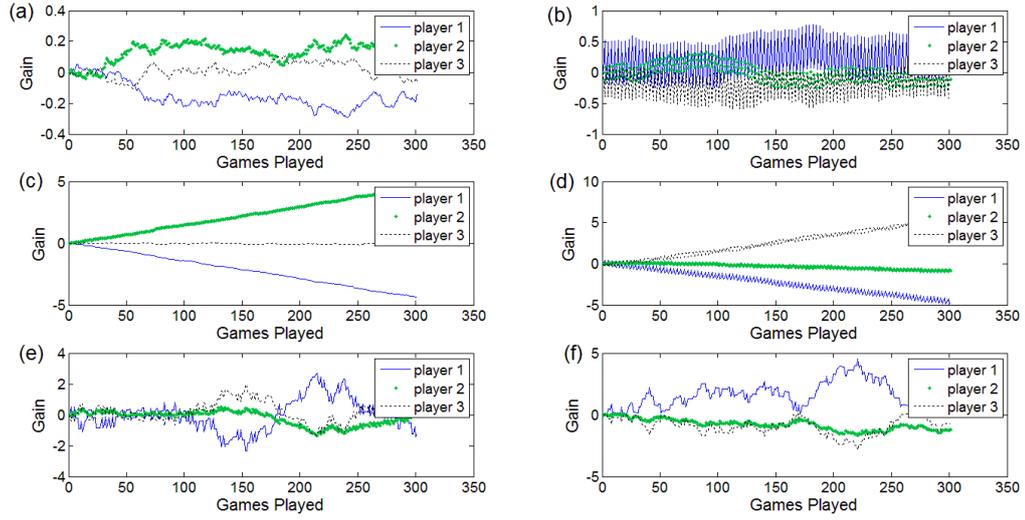

Figure 11. Individual games and random combination of them for $\varepsilon_1 = 0.005$, $\varepsilon_1 = 0.005$, and $\gamma = 0.5$. a) switch A before adding noise, b) switch B before adding noise, c) the effect of noise on switch A, d) the effect of noise on switch B, e) the effect of randomized switching between two fair switches A and B, f) the effect of randomized switching between two noisy switches A and B.

## 5. Discussion

In this work, we present a new individual version of Parrondo's games in which the two games, A and B, are played by three players instead of two players. We combine the games in random, periodic, and nonlinear orders. Unlike the periodic and nonlinear orders, the paradox does not hold for random order of the combination, but we can find new strategies of games that it holds. The existence of paradox can also be assayed for games that are played by more than three players. The described generalized model can be applied in several fields. One of its applications can be in genetic regulatory networks. We present three networks of new three-player games inspired by the possible regulatory relations among genes in genetic switches as a sample of genetic regulatory networks. There may not be a real specimen for these networks. However, this method can be a perfect starting point to design many other games for modeling different relations in regulatory networks and various diseases with drugs affecting them.

The deterministic conditions for switches provide the requirements for the occurrence of feedback loops. As a conditional game, Switch B is more robust to noise than switch A, which is a simple switch. This result can be obtained from the generalized model simulations and the games based on genetic switches. Previous researches confirm that feedback loops can enhance system stability in genetic switches [54-56, 63].

We assign each of the three statuses fair, winning, and losing to one of reactive elements. Suppose three reactive elements are in fair status. In that case, it can be inferred that the switch is in an unstable state in which any external signal can direct it to a stable fate. It can also be interpreted as a stable switch, and its decisions are compatible with the environment. Network III can be an example for the second interpretation. In this situation, if one of the regulatory elements digress from its correct path, the combined result of regulatory elements can remain in a correct destination. The other two networks (Networks I and II) are examples for the first interpretation of fair status for reactive elements in which the winning status is incompatible with the external environment.

The mathematical modeling of biochemical systems affected by noise is performed using stochastic methods. The chemical master equation (CME) [64-66], the stochastic simulation algorithm (SSA) [38, 64, 67], $\tau$-leaping algorithm [68, 69], chemical Langevin equation (RRE) [70], and hybrid methods are typical stochastic methods for modeling of biochemical systems. A game-theoretic approach is better than these methods for dynamic analysis of systems compatible with their environment [71]. The problem with the proposed method for modeling the genetic switches is the time of occurrence of the next reaction. In realistic systems, the subsequent reaction's time of occurrence is probabilistic [64, 67]. It can be resolved by randomizing the rounds to play the game. The algorithm is fast because only some of the reactions are considered in each regulatory system state. This assumption comes from the impossibility of occurrence of all reactions in each state [62, 72, 73]. The reactions occur with constant and predetermined probabilities in each state, but they can be fired with different probabilities in agreement with the states by defining new conditions for games [68, 69, 74, 75].

We acquire the fair probabilities for games in the two networks using *the Mathematica* software and simulate them in *MATLAB*. The results of simulations confirm the correctness of these mathematical relations.

## 6. Conclusion

In this work, we presented a new generalized model of Parrondo's games in which three players play the two games A and B. Game A is a simple non-cooperative game, and game B is a capital-dependent and non-cooperative game. We obtain the required conditions for fair, losing, and winning games theoretically. After that, we combine two games randomly, periodically, and nonlinearly. It was investigated that the paradox holds for a combination of two noisy games by periodic and nonlinear switching. This means that new strategies of games can be found in which the paradox is confirmed for a random combination of them.

This generalized model can be useful in several fields, including genetic switches. In the fourth section of this work, we consider three networks of games inspired by plausible regulatory relations among genes. As was demonstrated, a combination of genetic switches can improve network resistance to noise and the network can remain in its desirable path for specified bound of noise. This method can be applied in different fields of biology, pharmacy, and medicine.

## 7. References

1. Feynman, R.P., R.B. Leighton, and M. Sands, *The feynman lectures on physics; vol. i.* American Journal of Physics, 1965. **33**(9): p. 750-752.

2. Toral, R., P. Amengual, and S. Mangioni, *Parrondo's games as a discrete ratchet.* Physica A: Statistical Mechanics and its Applications, 2003. **327**(1): p. 105-110.

3. Parrondo, J. and L. Dínis, *Brownian motion and gambling: from ratchets to paradoxical games.* Contemporary Physics, 2004. **45**(2): p. 147-157.

4. Barton, R., *Chaos and fractals.* The Mathematics Teacher, 1990. **83**(7): p. 524-529.

5. Harmer, G.P., et al. *Parrondo's paradoxical games and the discrete Brownian ratchet.* in *AIP Conference Proceedings*. 2000. AIP.

6. Amengual, P., et al. *Discrete–time ratchets, the Fokker–Planck equation and Parrondo's paradox.* in *Proceedings of the Royal Society of London A: Mathematical, Physical and Engineering Sciences.* 2004. The Royal Society.

7. Harmer, G.P., et al., *Brownian ratchets and Parrondo's games.* Chaos: An Interdisciplinary Journal of Nonlinear Science, 2001. **11**(3): p. 705-714.

8. Harmer, G.P. and D. Abbott, *Parrondo's paradox.* Statistical Science, 1999: p. 206-213.

9.  Moraal, H., *Counterintuitive behaviour in games based on spin models.* Journal of Physics A: Mathematical and General, 2000. **33**(23): p. L203.

10. Tang, T.W., A. Allison, and D. Abbott, *Parrondo's games with chaotic switching.* arXiv preprint cs/0404016, 2004.

11. Arena, P., et al., *Game theory and nonlinear dynamics: the Parrondo paradox case study.* Chaos, Solitons & Fractals, 2003. **17**(2): p. 545-555.

12. Abbott, D., *Developments in Parrondo's paradox*, in *Applications of nonlinear dynamics*. 2009, Springer. p. 307-321.

13. Harmer, G.P. and D. Abbott, *A review of Parrondo's paradox.* Fluctuation and Noise Letters, 2002. **2**(02): p. R71-R107.

14. Amengual, P., *Paradoxical games: a Physics point of view.* 2006.

15. TAHAMI, E. and M.H. GOLPAYGANI, *The Role of Irrational Number Switching Strategies in the Parrondo's Paradox Game.*

16. Ethier, S.N. and J. Lee, *Parrondo's paradox via redistribution of wealth.* Electron. J. Probab, 2012. **17**(20): p. 1-21.

17. Harmer, G.P., D. Abbott, and P.G. Taylor. *The paradox of Parrondo's games*. in *Proceedings of the Royal Society of London A: Mathematical, Physical and Engineering Sciences*. 2000. The Royal Society.

18. Shu, J.-J. and Q.-W. Wang, *Beyond Parrondo's paradox.* arXiv preprint arXiv:1403.5468, 2014.

19. Parrondo, J.M., G.P. Harmer, and D. Abbott, *New paradoxical games based on Brownian ratchets.* Physical Review Letters, 2000. **85**(24): p. 5226.

20. Mihailović, Z. and M. Rajković, *Cooperative Parrondo's games on a two-dimensional lattice.* Physica A: Statistical Mechanics and its Applications, 2006. **365**(1): p. 244-251.

21. Toral, R., *Cooperative Parrondo's games.* Fluctuation and Noise Letters, 2001. **1**(01): p. L7-L12.

22. MIHAILOVIĆ, Z. and M. RAJKOVIĆ, *One dimensional asynchronous cooperative Parrondo's games.* Fluctuation and Noise Letters, 2003. **3**(04): p. L389-L398.

23. Toyota, N., *Does Parrondo Paradox occur in Scale Free Networks?-A simple Consideration.* arXiv preprint arXiv:1204.5249, 2012.

24. Ye, Y., L. Wang, and N. Xie, *Parrondo's games based on complex networks and the paradoxical effect.* PloS one, 2013. **8**(7): p. e67924.

25. Toyota, N., *Second Parrondo's Paradox in Scale Free Networks.* arXiv preprint arXiv:1207.4911, 2012.

26. Bier, M., *Brownian ratchets in physics and biology.* Contemporary Physics, 1997. **38**(6): p. 371-379.

27. Flitney, A. and D. Abbott, *Quantum models of Parrondo's games.* Physica A: Statistical Mechanics and its Applications, 2003. **324**(1): p. 152-156.

28. Grimm, A., *Separation and collective phenomena of colloidal particles in Brownian ratchets.* 2010.

29. Abbott, D., et al. *The problem of detailed balance for the Feynman-Smoluchowski engine (FSE) and the multiple pawl paradox.* in *AIP Conference Proceedings.* 2000. AIP.

30. Abbott, D., *Asymmetry and disorder: A decade of Parrondo's paradox.* Fluctuation and Noise Letters, 2010. **9**(01): p. 129-156.

31. Peskin, C.S., G.M. Odell, and G.F. Oster, *Cellular motions and thermal fluctuations: the Brownian ratchet.* Biophysical journal, 1993. **65**(1): p. 316-324.

32. Iyengar, R. and R. Kohli, *Why Parrondo's paradox is irrelevant for utility theory, stock buying, and the emergence of life.* Complexity, 2003. **9**(1): p. 23-27.

33. Stjernberg, F., *Parrondo's paradox and epistemology—when bad things happen to good cognizers (and conversely).* Hommage Wlodek. Philosophical Papers Dedicated to Wlodek Rabinowicz, 2007.

34. Khandelwal, P., et al., *Solution structure and dynamics of GCN4 cognate DNA: NMR investigations.* Nucleic acids research, 2001. **29**(2): p. 499-505.

35. Reed, F.A., *Two-locus epistasis with sexually antagonistic selection: a genetic Parrondo's paradox.* Genetics, 2007. **176**(3): p. 1923-1929.

36. Huang, S., et al., *Bifurcation dynamics in lineage-commitment in bipotent progenitor cells.* Developmental biology, 2007. **305**(2): p. 695-713.

37. Tian, T. and K. Smith-Miles, *Mathematical modeling of GATA-switching for regulating the differentiation of hematopoietic stem cell.* BMC systems biology, 2014. **8**(1): p. S8.

38. Guru, E. and S. Chatterjee, *Study of Synthetic Biomolecular Network in Escherichia Coli.* International Journal of Biophysics, 2013. **3**(1): p. 38-50.

39. Lopez, D., H. Vlamakis, and R. Kolter, *Generation of multiple cell types in Bacillus subtilis.* FEMS microbiology reviews, 2009. **33**(1): p. 152-163.

40. Ghaffarizadeh, A., N.S. Flann, and G.J. Podgorski, *Multistable switches and their role in cellular differentiation networks.* BMC bioinformatics, 2014. **15**(7): p. S7.

41. Lyons, S.M., et al., *Loads bias genetic and signaling switches in synthetic and natural systems.* PLoS Comput Biol, 2014. **10**(3): p. e1003533.

42. Balázsi, G., A. van Oudenaarden, and J.J. Collins, *Cellular decision making and biological noise: from microbes to mammals.* Cell, 2011. **144**(6): p. 910-925.

43. Perkins, T.J. and P.S. Swain, *Strategies for cellular decision‐making.* Molecular systems biology, 2009. **5**(1): p. 326.

44. Foster, D.V., et al., *A model of sequential branching in hierarchical cell fate determination.* Journal of theoretical biology, 2009. **260**(4): p. 589-597.

45. Cinquin, O. and J. Demongeot, *High-dimensional switches and the modelling of cellular differentiation.* Journal of theoretical biology, 2005. **233**(3): p. 391-411.

46. Caravagna, G., G. Mauri, and A. d'Onofrio, *The interplay of intrinsic and extrinsic bounded noises in biomolecular networks.* PLoS One, 2013. **8**(2): p. e51174.

47. Dorri, F., et al., *Natural biased coin encoded in the genome determines cell strategy.* PloS one, 2014. **9**(8): p. e103569.

48. Harmer, G.P. and D. Abbott, *Game theory: Losing strategies can win by Parrondo's paradox.* Nature, 1999. **402**(6764): p. 864-864.

49. Karlin, S. and H.E. Taylor, *A second course in stochastic processes*. 1981: Elsevier.

50. Parker, T.S. and L. Chua, *Practical numerical algorithms for chaotic systems*. 2012: Springer Science & Business Media.

51. Peitgen, H.-O., H. Jürgens, and D. Saupe, *Chaos and fractals: new frontiers of science*. 2006: Springer Science & Business Media.

52. Bucolo, M., et al., *Does chaos work better than noise?* IEEE Circuits and Systems Magazine, 2002. **2**(3): p. 4-19.

53. Fotoohinasab, A., Fatemizadeh, E., Pezeshk, H. and Sadeghi, M., 2018. *Denoising of genetic switches based on Parrondo's paradox.* Physica A: Statistical Mechanics and its Applications, 493, pp.410-420.

54. Ferrell, J.E., *Bistability, bifurcations, and Waddington's epigenetic landscape.* Current biology, 2012. **22**(11): p. R458-R466.

55. Zhang, H., Y. Chen, and Y. Chen, *Noise propagation in gene regulation networks involving interlinked positive and negative feedback loops.* PLoS one, 2012. **7**(12): p. e51840.

56. Leite, M.C.A. and Y. Wang, *Multistability, oscillations and bifurcations in feedback loops.* Math Biosci Eng, 2010. **7**(1): p. 83-97.

57. Kittisopikul, M. and G.M. Süel, *Biological role of noise encoded in a genetic network motif.* Proceedings of the National Academy of Sciences, 2010. **107**(30): p. 13300-13305.

58. Khalil, A.S. and J.J. Collins, *Synthetic biology: applications come of age.* Nature Reviews Genetics, 2010. **11**(5): p. 367-379.

59. Gardner, T.S., C.R. Cantor, and J.J. Collins, *Construction of a genetic toggle switch in Escherichia coli.* Nature, 2000. **403**(6767): p. 339-342.

60. MacArthur, B.D., A. Ma'ayan, and I.R. Lemischka, *Systems biology of stem cell fate and cellular reprogramming.* Nature Reviews Molecular Cell Biology, 2009. **10**(10): p. 672-681.

61. Baron, M., et al., *Designing Synthetic Gene Networks Using Artificial Transcription Factors in Yeast.* iGEM World Championship Jamboree, Cambridge, MA, 2013.

62. Gillespie, D.T., *Stochastic simulation of chemical kinetics.* Annu. Rev. Phys. Chem., 2007. **58**: p. 35-55.

63. Kim, J.-R., Y. Yoon, and K.-H. Cho, *Coupled feedback loops form dynamic motifs of cellular networks.* Biophysical journal, 2008. **94**(2): p. 359-365.

64. Ullah, M. and O. Wolkenhauer, *Stochastic approaches for systems biology*. 2011: Springer Science & Business Media.

65. Chen, W.-Y., *Stochasticity and noise-induced transition of genetic toggle switch.* Journal of Uncertainty Analysis and Applications, 2014. **2**(1): p. 1.

66. Strasser, M., F.J. Theis, and C. Marr, *Stability and multiattractor dynamics of a toggle switch based on a two-stage model of stochastic gene expression.* Biophysical journal, 2012. **102**(1): p. 19-29.

67. Cai, X., *Exact stochastic simulation of coupled chemical reactions with delays.* The Journal of chemical physics, 2007. **126**(12): p. 124108.

68. Wieder, N., R.H. Fink, and F. von Wegner, *Exact and approximate stochastic simulation of intracellular calcium dynamics.* BioMed Research International, 2011. **2011**.

69. Ahn, T.-H., Y. Cao, and L.T. Watson. *Stochastic Simulation Algorithms for Chemical Reactions*. in *BIOCOMP*. 2008.

70. Ilie, S., W.H. Enright, and K.R. Jackson, *Numerical solution of stochastic models of biochemical kinetics.* Canadian Applied Mathematics Quarterly, 2009. **17**(3): p. 523-554.

71. Pfeiffer, T. and S. Schuster, *Game-theoretical approaches to studying the evolution of biochemical systems.* Trends in biochemical sciences, 2005. **30**(1): p. 20-25.

72. El Samad, H., et al., *Stochastic modelling of gene regulatory networks.* International Journal of Robust and Nonlinear Control, 2005. **15**(15): p. 691-711.

73. Ribeiro, A.S., *Stochastic and delayed stochastic models of gene expression and regulation.* Mathematical Biosciences, 2010. **223**(1): p. 1-11.

74. De Jong, H., *Modeling and simulation of genetic regulatory systems: a literature review.* Journal of computational biology, 2002. **9**(1): p. 67-103.

75. Tian, T. and K. Burrage, *Stochastic models for regulatory networks of the genetic toggle switch.* Proceedings of the national Academy of Sciences, 2006. **103**(22): p. 8372-8377.